\documentclass{aa501}
\usepackage{graphicx}
\begin{document}
%
%

\title{RXTE observations of single pulses of PSR B0531+21 I: Flux variations}

\titlerunning{RXTE observations of Crab pulsar}

\author{M. Vivekanand\thanks{vivek@ncra.tifr.res.in}}

\institute{National Center for Radio Astrophysics, TIFR, Pune University Campus,
P. O. Box 3, Ganeshkhind, Pune 411007, India.}

\date{Received (date) / Accepted (date)}

\abstract{
This article is the first in the series that analyze about 1.87 million periods 
of PSR B0531+21 (Crab pulsar), observed by the PCA detector aboard the RXTE 
x-ray observatory.  The Crab pulsar's x-ray light curve shows little variation 
over time scales ranging from days to a period (33.46 milli seconds). The 
standard deviation of its x-ray flux variation is $\approx$ 0.7\% of its mean 
value, which is negligible compared to its radio flux variations. The phase 
resolved power spectrum of pulse to pulse x-ray flux variation shows no spectral 
feature; an upper limit to the peak of any possible broad spectral feature is 
0.06\% of the mean power. The x-ray fluxes in the two components of its 
integrated profile are unrelated to each other; their linear correlation 
coefficient is 0.0004$\pm$0.0010.  ``Giant pulses'' that are routinely seen at 
radio wavelengths are absent here. This work sets very strong constraints on 
the connection (if any) between the flux variations at radio and x-ray energies,
for example due to variation in the degree of coherence of the basic emitters.
Its phase resolved x-ray flux variation shows a weak correlation with 
the integrated profile. If confirmed, this might be an important clue to 
understanding the x-ray emission mechanism of Crab pulsar.
\keywords{pulsars -- PSR B0531+21 -- single pulses -- x-ray -- RXTE.}
}

\maketitle

\section{Introduction}

The intrinsic temporal flux variations of pulsars are probably an 
important clue to 
understanding their emission mechanism at any wavelength. In fact, the 
flux variations at high energies (optical, x-ray, $\gamma$-ray) might 
provide important constraints to the emission mechanism at radio 
wavelengths also (Cheng et al, \cite{CHR2}; Kawai et al \cite{KOB91};
Lundgren et al \cite{LCM95}; 
Moffet \& Hankins \cite{MH1996}; Patt el al \cite{PUZ99}). Patt el al
(\cite{PUZ99}) searched for period to period flux variations in about 
105\,000 periods of Crab pulsar data at x-ray energies (in the range 1 
to 10 kilo electron volts (KeV)), with a time resolution of 100 micro 
seconds ($\mu$s), obtained by the PCA detector aboard the RXTE x-ray 
observatory. This work reports the results of analyzing 1\,868\,112 
periods of Crab pulsar from the same instrument, with a time resolution 
of 3.815 $\mu$s, in the energy range 13.3 to 58.4 KeV.

The RXTE data archive was searched using the XDF tool, for public data 
acquired by the Proportional Counter Array (PCA). A uniform set of 23 data 
files were found observed during August/September 1996, their ObsId numbers 
ranging from 10203-01-01-00 to 10203-01-03-01. They were obtained in the EVENT 
mode (XTE\_SE), combining events from all five Proportional Counter Units 
(D[0\symbol{126}4]), and also from both halves of all three Xenon anode layers 
of each PCU (X1L\symbol{94}X1R\symbol{94}X2L\symbol{94}X2R\symbol{94}X3L\symbol{94}X3R).
Channels 50 to 249 of the PCA were also combined, which corresponds to the 
energy range 13.3 to 58.4 KeV.

The first phase of data analysis used the FTOOLS software. First, 
the Good Time Intervals (GTI) were obtained for each data file by using the 
MAKETIME tool on the corresponding XTE filter file; the selection criterion 
were (a) pointing OFFSET less than 0.02$\degr$, (b) elevation (ELV) greater 
than 10$\degr$, and (c) all five PCUs to be switched on. Next, the GTI 
extension of each data file was edited to insert the above GTI values. Then 
the FSELECT tool was run to filter out data outside these time ranges. Next 
the SEFILTER tool was run with the M[1]\{1\} option (without bypassing the 
FSELECT tool) to retain only the valid photon events. Then the FXBARY tool 
was run using the orbit file for that day, to convert the arrival times of 
photons from the Terrestrial Time system (TT) to the solar system 
barycenter system (TDB). Penultimately, the SEEXTRCT tool was used 
to obtain the light curve for each file, in time intervals of 1.010895 milli 
second (ms), which is 265 times the basic time resolution of the data. 
Finally, the PCADTLC tool was used to correct the light curve for dead time 
of the PCA; before this the corresponding Standard 1 files were also converted 
to the TDB system for consistency. 

Time samples having incomplete exposure were deleted.  These occurred naturally 
at the beginning and end of each light curve, and also whenever the RXTE 
observatory shut off some PCUs, for technical reasons.  Each light curve was 
then converted into the ASCII format for further processing. 

The second phase of data analysis used self-developed software. First the 
power spectrum of each light curve was computed to obtain the period of Crab 
pulsar in that file. This was used to separate the light curve into individual 
periods (also called single pulses). Each period has 33 time samples (also 
called bins), giving a synthesized sampling interval that is different from 
file to file, but is $\approx$ 1.013967 ms. The above separation was done such 
that the photon counts in each original sampling interval were not split across 
more than one synthesized sampling intervals; otherwise the Poisson statistics 
of the data would be distorted, and would cause problems for some studies as 
discussed later on. The separation into individual periods for radio data is 
much simpler, since standard resampling techniques can be used. 

\begin{figure}
\resizebox{\hsize}{!}{\includegraphics{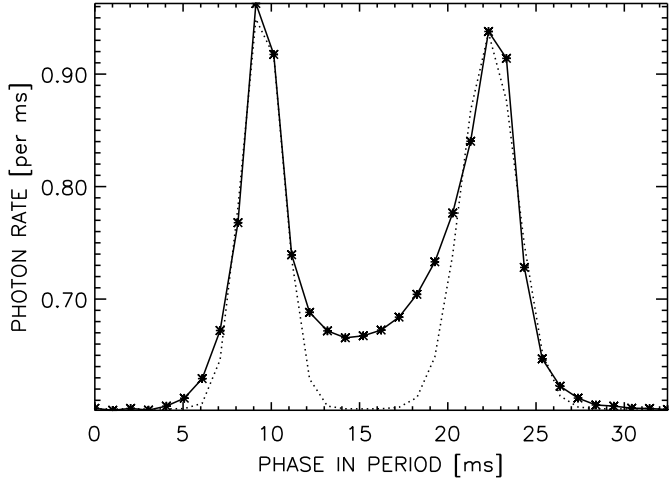}}
\caption{Integrated profile of Crab pulsar after summing 1\,868\,112 periods 
         from 23 files. The abscissa is time (also called 
	 phase) within the period (in ms), while the ordinate is the average 
	 number of photons obtained in one synthesized time sample (1.013967 ms).
	 The mis-alignment of the profiles from file to file does not exceed 
	 half a synthesized time sample. The dotted curves represent the two 
	 peaks modeled as Gaussian.}
\label{fig1}
\end{figure}

The appropriate period for each data file was ascertained more accurately by 
checking for ``drift'' of the integrated profile between the first and second 
halves of the light curve; this part was done iteratively (see Vivekanand et al 
\cite{VAM98} for details). A straight line fit, to the starting epoch (TDB 
system) of each data file versus the Crab pulsar period in that file, gives a 
period derivative of 4.208 ($\pm$ 0.005) $\times 10^{-13}$ s/s, which compares 
excellently with the actual value for Crab pulsar. The standard deviation of 
the periods about the fitted straight line is $\approx$ 2 nano seconds (ns), 
which is consistent with the expected value; the data files contain typically 
50\,000 to 98\,000 periods, and one can recognize a relative shift of a small 
fraction of the time sample between the integrated profiles of the first and 
second halves of a data file. However the periods differ systematically by 
$\approx$ 8 ns from those obtained using Crab pulsar's ephemeris. It is not 
clear to this author why this systematic difference should occur, but this does 
not affect the rest of the analysis.

Fig.~\ref{fig1} shows the integrated profile of Crab pulsar for 1\,868\,112 
periods. The integrated profiles of all files are aligned, correct to half a 
time sample, so that one can analyze them as if all 23 files have been obtained 
``in phase''. However it does smear the integrated profile to a maximum 
of half a synthesized time sample. Samples 5 to 30 are considered to represent 
the {\it on-pulse} window, and the rest of the seven samples the {\it off-pulse} 
window, although the Crab pulsar might emit x-rays all through its period.

Details of the analysis in the coming sections can be found in Vivekanand \& Joshi 
\cite{VJ97}, Vivekanand et al \cite{VAM98}, and Vivekanand \cite{MV2000}; they will
be described only briefly in this article.

\section{X-ray light curve}

\begin{figure}
\resizebox{\hsize}{!}{\includegraphics{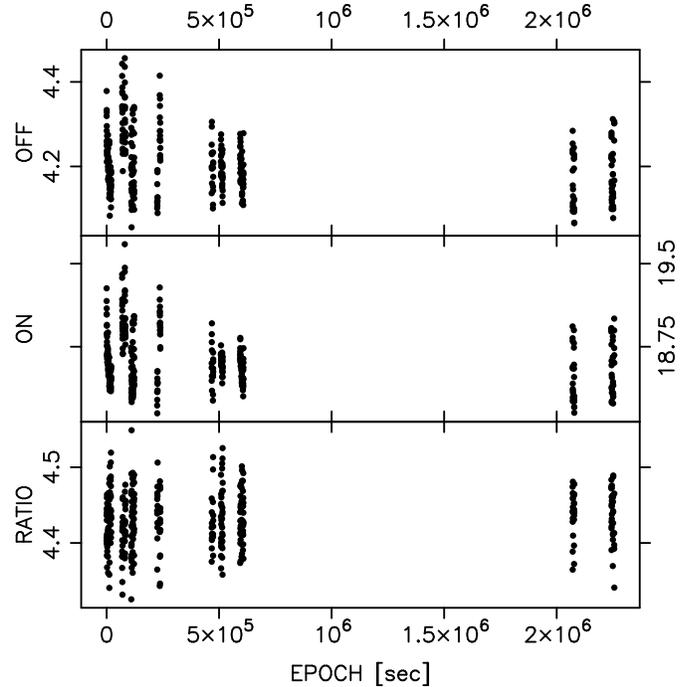}}
\caption{X-ray light curve of Crab pulsar over 26 days during August/September
         1996. The first two plots show the average number of photons in the 
         off-pulse window (OFF) and on-pulse window (ON) defined in Fig. 
	 ~\ref{fig1}. The last plot is the ratio ON/OFF.}
\label{fig2}
\end{figure}

Figure~\ref{fig2} shows the average number of photons in the off- and on- 
pulse windows of Fig.  ~\ref{fig1}. Each point in the figure represents the 
average number of photons in 5\,000 periods ($\approx$ 167 s); the total 
duration in the figure is 26.1 days. Both ON and OFF counts show correlated 
hourly as well as daily variation. Assuming that this is most probably an 
instrumental effect, the ratio ON/OFF in Fig.~\ref{fig2} would represent 
the true variation of the Crab pulsar x-ray flux. The mean value of the ratio 
is 4.432 while its standard deviation is 0.035; so the x-ray flux of Crab pulsar 
fluctuates with a standard deviation of 0.035/4.432 $\approx$ 0.8\% of its 
mean value, which is a negligible quantity. This supports the reported 
Crab pulsar behavior at 0.015 to 0.130 mega electron volt (MeV) energies 
(Ubertini et al \cite{UBC94}), but is contrary to its behavior at $\approx$ 
200 MeV reported by Nolan et al (\cite{NOLAN93}) using CGRO observations, 
where they found the $\gamma$ ray flux of Crab pulsar to vary over time 
scales of a month. However the above result (0.8\%) would be an under 
estimate if there is significant emission from the Crab pulsar during the 
off-pulse window.

It would have been interesting to study the existence, at x-rays, of the 
reported very long time scale (of $\approx$ 13.5 years) variation of the 
ratio of the fluxes in the two peaks (of Fig.~\ref{fig1}) at $\gamma$-rays 
(Nolan et al \cite{NOLAN93}); unfortunately this can not be done due to 
the limited data span available here.

\section{Pulse to pulse flux variations}

The last section discussed the flux variations of Crab pulsar over time scales 
of hours and days. The current section discusses flux variations from period 
to period. 

\subsection{Spectrum of flux variations}

\begin{figure}
\resizebox{\hsize}{!}{\includegraphics{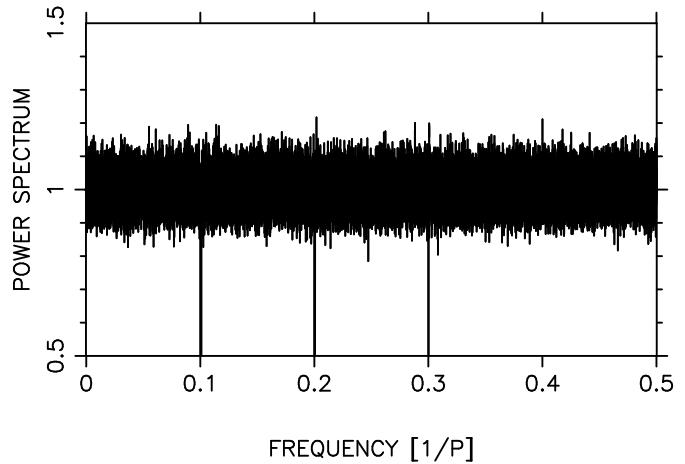}}
\caption{Average power spectrum of pulse to pulse x-ray flux variations of 
	 Crab pulsar. The time series, consisting of x-ray photons as a 
	 function of the period number, was Fourier transformed at each of 
	 the 33 phases of the integrated profile; the modulus squared of 
	 the 33 transforms was averaged. The abscissa in is units of inverse 
	 periods, extending up to 0.5 due to the Nyquist criterion. A small 
	 range of frequencies has been removed around the sampling spectral 
	 feature ($\approx$ 1/10 periods), and two of its harmonics.}
\label{fig3}
\end{figure}

Figure~\ref{fig3} shows the so 
called x-ray fluctuation spectrum of Crab pulsar. At each of the 33 samples 
(phases) of the integrated profile of Fig.~\ref{fig1}, a time series was formed 
comprising of the x-ray flux as a function of the period number in the data file. 
This was Fourier transformed in arrays of length 128 $\times$ 1024 periods. The 
data were centered in the array and zero padded, and then a Hamming window was 
applied. To remove long term variations, the data in blocks of 32\,768 periods were 
normalized with the mean value of this block (see Ritchings \cite{RT76} and 
Vivekanand \& Joshi \cite{VJ97} for details). Fourteen data files were chosen that 
had at least 75\,000 periods each, totalling to 1\,347\,028 periods. 
Figure~\ref{fig3} shows the power spectrum averaged over the 33 spectra,
after normalizing each spectrum with the variance of its time series.
A polynomial of the form

\begin{equation}
y = a_0 + a_1 x + a_2 x^2 + a_3 x^3
\end{equation}

\noindent
was fit to the power spectrum in Fig.~\ref{fig3}. the coefficients are $a_0 =
0.999 \pm 0.001$, $a_1 = 0.003 \pm 0.013$, $a_2 = -0.003 \pm 0.059$ and $a_3 = 
-0.004 \pm 0.077$. The standard deviation of the power spectrum with respect
to the above fitted curve is 0.046, which is mainly determined by photon noise. 
It is clear that none of the coefficients are significant except the first. 

The upper limit to any possible weak and broad spectral feature that might be 
hidden in the data can be computed to be $\approx$ 0.06\% of the total power in
the spectrum. Therefore Fig.~\ref{fig3} is consistent with the Crab pulsar having 
no spectral feature in its pulse to pulse x-ray flux variations.

\subsection{Modulation index}

The radio flux of rotation powered pulsars varies significantly from pulse to
pulse; this could be due to intrinsic flux variation of the sub pulse, as
well as random position of the sub pulse within the on-pulse window (here one
is ignoring the flux variations due to propagation in the interstellar medium). 
This is characterized by the so called modulation index $\mu$, defined as

\begin{equation}
\mu = \frac{\sigma_I}{\langle I \rangle}
\end{equation}

\noindent
where $\langle I \rangle$ and $\sigma_I$ are the mean and the standard deviation of 
the pulsar flux (see Manchester \& Taylor \cite{MT77}); $\mu$ represents the 
fractional flux variation of the pulsar, and is usually greater than 1.0 at 
radio wavelengths.

\begin{figure}
\resizebox{\hsize}{!}{\includegraphics{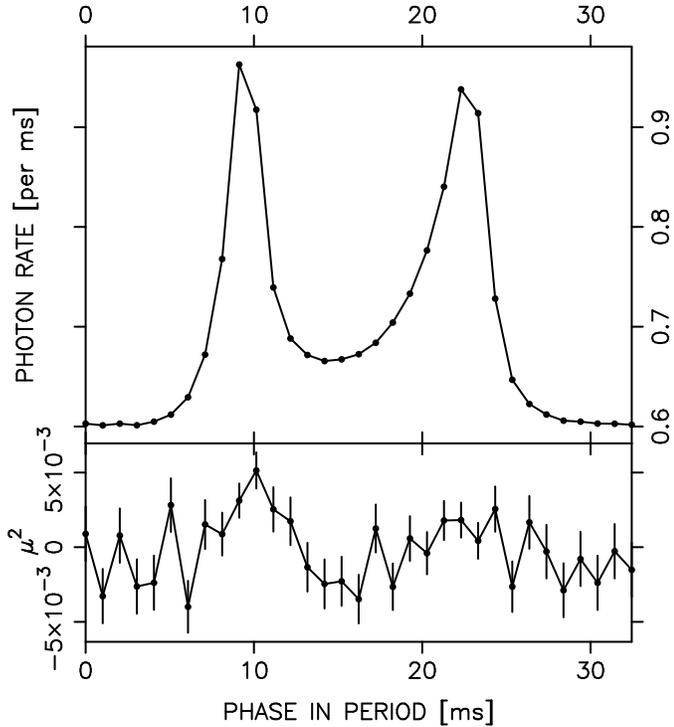}}
\caption{Square of the estimated modulation index ($\mu^2$) of the Crab pulsar 
	 x-ray flux, at each sample of the integrated profile. The top panel 
	 is the same as Fig.~\ref{fig1}. The vertical bars represent two 
	 standard deviation error bars. Since $\mu^2$ depends upon the small 
	 difference between two much larger quantities (eq. 3), it can be
	 negative also depending upon measurement errors. The distribution
	 of these negative values about the value 0.0 is consistent with 
	 their rms errors, as expected.}
\label{fig4}
\end{figure}

Figure~\ref{fig4} shows the square of $\mu$, which is the natural quantity
to average, for all 1\,868\,112 periods. First, the mean $\langle I \rangle$ and the 
variance $\sigma^2$ are computed at each sample in the integrated profile ($\langle 
I \rangle$ is plotted in Fig.~\ref{fig1} and the top panel of Fig. 
~\ref{fig4}). Now $\sigma^2$ has contribution from two sources:

\begin{equation}
\begin{array}{ll}
\sigma^2 & = \sigma_K^2 + \sigma_I^2 \\
         & = \langle I \rangle  + \sigma_I^2 \\
\end{array}
\end{equation}

\noindent
where $\sigma_K^2$ is the variance due to Poisson statistics of photons, and 
equals the mean number of photons $\langle I \rangle$, while $\sigma_I^2$ represents the
fluctuation of the average intensity of the pulsar; the two are referred to 
as photon noise and wave noise, respectively (Goodman \cite{JG85}). The
modulation index squared in Fig.~\ref{fig4} was estimated by subtracting the mean 
intensity $\langle I \rangle$ from the estimated variance $\sigma^2$, and then dividing 
by $\langle I \rangle^2$ at each sample of the integrated profile. Because of the PCA 
dead time correction to the data, the photon rate of Crab pulsar in any 
time sample is about 6\% higher than the corresponding integer value of 
photons. This correction was estimated self-consistently by averaging $\sigma^2 
- \langle I \rangle$ over all 33 samples of the integrated profile, for each file. The 
average value of the correction for all 23 files is 1.062 with standard 
deviation of 0.003. The mean flux $\langle I \rangle$ at each sample was scaled by this 
constant (for that data file) before subtracting from $\sigma^2$. This constant is
not dependent upon the average flux $\langle I \rangle$ at each sample, so 
the above procedure is unlikely to introduce artifacts in the $\mu^2$ of Fig. 
~\ref{fig4}.

The average value of $\mu^2$ in Fig.~\ref{fig4} is $-0.0002 \pm 0.0040$, which
is is consistent with zero as expected. The $\chi^2$ of $\mu^2$ of Fig. 
~\ref{fig4} with respect to the expected value of 0.0 is 76.3 for 33 degrees of 
freedom. Removing sample number 10 reduces the $\chi^2$ to 57.0, and further 
removing sample number 11 reduces it to 48.7, which is just 2.25 standard 
deviations away from the expected value of 31.0. Therefore to the zeroth order 
of accuracy it is reasonable to assume that $\mu^2$ is the same (i.e., 0.0) for 
all samples in the integrated profile. Its standard deviation is 0.0024 in the 
on-pulse window; then the standard deviation of $\mu$ can be estimated as 
$\sqrt{0.0024 / 2} \approx$ 3.5\% (by simple algebra). Thus the rms x-ray flux 
variation at any phase in the on-pulse window of the integrated profile of Crab 
pulsar is $\approx$ 3.5\% of its mean value. Then the rms variation of the total 
on-pulse x-ray flux of the Crab pulsar will be $\approx$ 3.5 / $\sqrt{26} 
\approx$ 0.7\%. A similar calculation for the entire integrated profile gives 
$\sqrt{0.0023 / 2} / \sqrt{33} \approx$ 0.6\%, which is essentially the same 
result. This is a much tighter limit than the $\approx$ 7\% quoted by Patt el al 
(\cite{PUZ99}). These authors used totally 105\,000 periods and a different 
method of analysis, on account of which their result might be dominated by 
photon noise. The result of this section should ideally reflect the actual 
x-ray intensity variations of Crab pulsar (wave noise), that contain 
information about the physics of the x-ray emission mechanism.

To the next order of accuracy $\mu^2$ appears to be correlated with the shape 
of the integrated profile; both the lower panel of Fig.~\ref{fig4} as
well as the $\chi^2$ discussed above point to this. The $\mu^2$ at the phase 
of the second peak of the integrated profile also appears to be enhanced. The
on-pulse and off-pulse $\chi^2$ are 69.0 and 7.3, for 26 and 7 degrees of
freedom respectively; the former is 6 standard deviations away from the
expected value, which is quite high. However more data, or better analysis, is 
needed to confirm this with good statistical significance.

\subsection{Giant pulses}

Comparison of photon counts in the off- and on-pulse windows of Fig.~\ref{fig1} 
shows that the 
Crab pulsar does not emit ``giant pulses'' at x-ray energies. The mean on-pulse 
window photon rate is 17.57 photons in 26 time samples, while the maximum is 
about 31 photons. This implies that at x-ray energies the Crab pulsar emission 
occasionally increases by $\approx$ 31.0 / 17.57 $\approx$ 1.76 times, at the 
most, whereas at radio wavelengths the giant pulse energy is about 10 to 100 
times its mean value (Lundgren et al \cite{LCM95}). Further, the on-pulse photon
distribution fits a Poisson distribution very well, and there is no discernible 
excess probability at higher photon rates. Finally, following the method of 
Ritchings (\cite{RT76}) and Vivekanand (\cite{MV95}), a deconvolved photon 
distribution was obtained that represents the true on-pulse photon distribution 
of the Crab pulsar; it also does not indicate the presence of an excess 
probability at higher photon rates.

\section{Correlation of flux in the two peaks}

Patt et al (\cite{PUZ99}) report that the x-ray flux in the two peaks of the
integrated profile of Fig.~\ref{fig1} are uncorrelated. Their integrated profile
consists of 16 samples, in which the two peaks are defined as covering phases 1 
to 4 plus 15 and 16, and phases 5 to 11, respectively, (Fig. 1 of Patt et al 
\cite{PUZ99}). They obtain the covariance between the number of photons in these 
two peaks to be $-7.7 \times 10^{-4}$, which they claim is consistent with zero.
In this work the dotted curves of Fig.~\ref{fig1} are the models for the two peaks. 
The x-ray flux in each period was multiplied by the two curves, to obtain the flux 
in the two peaks, respectively. A normalized correlation coefficient (in contrast
to the covariance of Patt et al \cite{PUZ99}) was computed for these two fluxes, 
across all 1\,868\,112 periods; the result is 3.5 ($\pm$ 10.0) $\times$ 10$^{-4}$, 
which is consistent with zero. Therefore it is concluded that the x-ray fluxes of 
Crab pulsar in the two peaks of its integrated profile are indeed uncorrelated.

\section{Discussion}

The x-ray flux variations of Crab pulsar have a standard deviation of $\approx$ 
0.6\% to 0.8\% of its mean value over time scales ranging from a period to almost
a month; that this number is similar over such wide time scales may or may not
be a coincidence. This is consistent with the Crab pulsar's behavior at optical 
and UV energies (Percival et al \cite{PBD93}) and at IR energies (Lundgren et al 
\cite{LCM95}). Therefore the Crab pulsar flux variations at higher energies are
insignificant compared to those at radio wavelengths, where the modulation index 
$\mu$ is $\approx$ 1.0, most of which apparently comes from the giant pulses 
(Lundgren et al \cite{LCM95}). Now, this causes problems for the conjecture that 
both the radio and the high energy emission are related through a common electric 
current. If the 
flux variations of Crab pulsar are due to temporal variation in the number of 
basic emitters, then the radio flux variations would also have been at the 
$\approx$ 0.7\% level. If they are due to temporal variation in the coherence of 
the basic emitters, then the radio flux variations would have been at the 2 
$\times$ 0.7 $\approx$ 1.4\% level, since the intensity of a coherent emission 
mechanism is proportional to the square of the number of basic emitters. If they 
are due to temporal variation in the angle of some elementary beams (Lundgren et al 
\cite{LCM95}), one would have expected the radio flux variations to be smaller
than those at high energy, since the common beams would be larger at the larger 
wavelengths. One could surely postulate variations in the angle of radio beams
only, but that would have to be justified on the basis of some other independent 
physical mechanism. In summary the difference in the radio and high energy flux 
variations of Crab pulsar is difficult to explain, if the basic charged emitters
at both wavelengths are somehow related.

Another method, of amplifying the very small flux variations of Crab pulsar at 
high energies to the very large variations at radio wavelengths, would be
to somehow use a fraction of the $\approx$ 10$^7$ amplification factor of 
particles in the gaps, due to cascading e$^+$--\,e$^-$ pair production (Ruderman 
\& Sutherland \cite{RS75}; Cheng et al \cite{CHR1}, \cite{CHR2}).
One could probably postulate that the radiation that is emitted by the later 
generation of charges in the pair cascade process suffers greater variation in 
its intensity, due to amplification of the variation in the number density of
charges. This might also imply that one should see a monotonic increase in the
modulation index as one observes at larger wavelengths. More detailed study of 
Cheng et al (\cite{CHR1}, \cite{CHR2}) and Romani \& Yadigaroglu (\cite{RY95})
models is required for a reasonable solution. In any case, the 
results of this paper set very strong constraints on the explanation for the 
relative flux variations at the radio and x-ray energies.

The possible correlation of $\mu$ with the integrated profile in Fig.~\ref{fig4},
if confirmed in future, might set very strong constraints on the basic emission
mechanism of high energy emission from the Crab pulsar. In the framework of Cheng 
et al (\cite{CHR1}, \cite{CHR2}) and Romani \& Yadigaroglu (\cite{RY95}) models, 
the two peaks of the integrated profile 
are cusps created by emission from different magnetic field lines, that add in 
phase along different directions due to relativistic aberration. A well defined 
relation between the mean flux and its variance at each point in the integrated 
profile (for example, $\mu$ might vary as $\langle I \rangle^\alpha$) will be an additional 
constraint, along with the exact shape $\langle I \rangle$ of the integrated profile, on 
the x-ray emission mechanism.

Further work on these data is in progress, that studies issues such as (a) 
verifying if the Crab pulsar shows at x-ray energies the three phenomenon that 
are often seen in several radio pulsars -- ``pulse nulling'' ``systematic sub 
pulse drifting'' and ``mode changing''; (b) looking for special behaviour in
the x-ray integrated profile at the phase of the radio precursor, which is 
supposed to be different from the rest of the radio integrated profile; (c)
comparison of the peak and the bridge x-ray emission of the Crab pulsar,
which might further constrain the models of Cheng et al (\cite{CHR1}, 
\cite{CHR2}), Romani \& Yadigaroglu (\cite{RY95}), and Cheng et al 
\cite{CRZ00}.

On the theoretical front, it is probably worth exploring the simultaneous 
modeling of $\langle I \rangle$ and $\sigma^2_I$ (or $\mu$) at x-rays for rotation
powered pulsars, and more specifically for the Crab pulsar.

\begin{acknowledgements}

This research has made use of (a) High Energy Astrophysics Science Archive 
Research Center's (HEASARC) facilities such as their public data archive, and 
their FTOOLS software, and (b) NASA's Astrophysics Data System (ADS) 
Bibliographic Services. The author is thankful to them for their excellent
services.

\end{acknowledgements}

\end{document}